\documentclass[12pt,a4paper]{article}
\usepackage{mathtext}
\usepackage{graphicx}
\begin{document}
\author{B.~Pavlyshenko   \\
  \small  Ivan Franko Lviv National University (electronics department),
\\ \small Dragomanov Str. 50, Lviv, 79005 Ukraine.  \\ \small e-mail:
pavlsh@yahoo.com  }

\title{Quantum Algorithm of Evolutionary Analysis of 1D
 Cellular Automata}
\date{}
\maketitle
\begin{abstract}

It is shown that irreversible classical cellular automata can be
performed by quantum algorithm using additional ancilla registers.
The algorithm for cellular automata states analysis has been
proposed. This algorithm is based on the elements of Grover's
algorithm - the inversion of amplitude of searched states and
unitary transform of inversion about the average. The  inversion
of searched states amplitudes can be performed by quantum Toffoli
gate.
\end{abstract}

\section*{Introduction}

The cellular automata (CA) is a computational model which has been
studied in many scientific works . These models are based on
simple updating rules and show the perspective for computational
applications.

The aim of our study is to design the algorithms which allow to
use quantum parallelism for the  investigation of CA evolution for
all initial states simultaneously. For example, if one dimensional
CA consists from only 100 cells with 2 possible states, then one
quantum evolution can evaluate $2^{100}$ evolutions for the same
number of initial states. It is important to answer a question -
does or does not   some searched finish state ("good state") in CA
evolution for analyzed CA updating rules exist?  We will use
Grover's algorithms ideas to answer this question in  case of one
dimensional quantum cellular automata.

Quantum cellular automata (QCA) are being investigated in many
modern works. In the work \cite{qca1}   QCA for universal quantum
computation has been presented. In \cite{qca2,qca8}   one
dimensional QCA has been investigated. In \cite{qca3}   QCA
formalism based on a lattice of qubits has been described. In
\cite{qca4,qca5}   reversible QCA was considered. In
\cite{qca6,qca7}  the survey of QCA investigations has been
presented.

\section*{One Dimension Cellular Automata }

Let us consider the ability of implementation of synchronous
cellular automata by the quantum logical gates when all cells
switch to a new state simultaneously. Such a circuit can be
performed by using additinal qubits named ancilla. Let us consider
for determinance one of CA rules for updating CA states. For
example, a cell switches to state "0" if its   neighbors have the
same value, otherwise this cell switches to value "1". This rule
can be written as
\begin{equation}
x^{'}[i]:=x[i-1]\oplus x[i+1] \label{eq9}.
\end{equation}
In the classical CA such evolution is irreversible. If the results
of previous iteration are saved in an additional ancilla, then CA
evolution reversibility and unitarity can be achieved. Consider
unitary operator for performing rule (\ref{eq9}).
 This operator acts on three qubits - $|x_{i-1}\rangle,|x_{i+1}\rangle$ ,  $|a_{i}\rangle$
two neighbors  and ancilla from the additional register:
\begin{equation}
C=(X\otimes X \otimes I) \cdot T \cdot (X\otimes X\otimes I) \cdot
T \cdot( I\otimes I \otimes X) \label{eq10}.
\end{equation}
Before the iteration all additional ancilla are in the state
\begin{equation}
|0,0,\ldots,0\rangle_{n}
\end{equation}
The series of unitary transformations for qubits can be written as
\begin{eqnarray}
C_i:  I\otimes I \otimes X |x_{i-1},x_{i+1},0 \rangle \rightarrow
T|x_{i-1},x_{i+1},1 \rangle \rightarrow |x_{i-1},x_{i+1},1\oplus
x_{i-1}x_{i+1} \rangle
\nonumber \\
\rightarrow   X \otimes X \otimes I |x_{i-1},x_{i+1},1\oplus
x_{i-1}x_{i+1} \rangle \rightarrow |\neg x_{i-1},\neg
x_{i+1},1\oplus x_{i-1}x_{i+1} \rangle  \nonumber \\
\rightarrow T|\neg x_{i-1},\neg x_{i+1},1\oplus x_{i-1}x_{i+1}
\rangle \rightarrow |\neg x_{i-1}, \neg x_{i+1},1\oplus
x_{i-1}x_{i+1} \oplus \neg x_{i-1} \neg x_{i+1}  \rangle \nonumber \\
\rightarrow X \otimes X \otimes I |\neg x_{i-1}, \neg
x_{i+1},1\oplus x_{i-1}x_{i+1} \oplus \neg x_{i-1} \neg x_{i+1}
\rangle \rightarrow |x_{i-1},x_{i+1}, x_{i}^{t+1} \rangle
\end{eqnarray}
The action of operator  $C_i$ on the possible qubits values can be
described as
\begin{eqnarray}
C_i:|0\rangle_{i-1,t}|0\rangle_{i+1,t}|0\rangle_{a,t+1}\rightarrow
|0\rangle_{i-1,t}|0\rangle_{i+1,t}|0\rangle_{a,t+1} \nonumber \\
C_i:|1\rangle_{i-1,t}|1\rangle_{i+1,t}|0\rangle_{a,t+1}\rightarrow
|1\rangle_{i-1,t}|1\rangle_{i+1,t}|0\rangle_{a,t+1} \nonumber \\
C_i:|1\rangle_{i-1,t}|0\rangle_{i+1,t}|0\rangle_{a,t+1}\rightarrow
|1\rangle_{i-1,t}|0\rangle_{i+1,t}|1\rangle_{a,t+1} \nonumber \\
C_i:|0\rangle_{i-1,t}|1\rangle_{i+1,t}|0\rangle_{a,t+1}\rightarrow
|0\rangle_{i-1,t}|1\rangle_{i+1,t}|1\rangle_{a,t+1} \label{eq13}
\end{eqnarray}

\begin{figure}[h]
\centering
\includegraphics[width=0.5\textwidth]{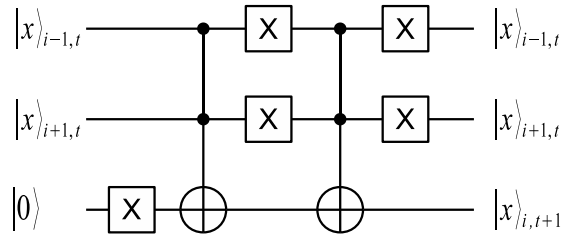}
\caption {{\small  Quantum circuit implementation for CA  updating
rules }} \label{fig1}
\end{figure}

 CA  updating rules (\ref{eq9})-(\ref{eq13}) can be described by quantum circuit in
fig.\ref{fig1}

CA state at $ j $ iteration can be described using the following
operator
\begin{equation}
A_j=C_n \ldots C_2 C_1 \label{eq14}.
\end{equation}
Let us consider the first CA iteration with the additional ancilla
register.
\begin{equation}
A_1: | x \rangle_0 | 0 \rangle_1 \rightarrow | x \rangle_0 | f(x)
\rangle_1 \label{eq15}.
\end{equation}
where $|x\rangle = |x_1,x_2 \ldots x_n \rangle $ - is initial CA
register, $ f(x) $ - is some function which expresses CA rules for
CA updating (\ref{eq9})-(\ref{eq13}). On the next iteration the
new ancilla register is being added. It can be described as
\begin{equation}
A_2: | x \rangle_0 | f(x) \rangle_1 | 0 \rangle_2^{\otimes n}
\rightarrow | x \rangle_0 | f(x) \rangle_1 | f(f(x)) \rangle_2
\label{eq16}.
\end{equation}
CA evolution after $ m $ iteration can be written as
\begin{equation}
A_m \ldots A_2 A_1: | x \rangle_0 | 0 \rangle_1^{\otimes n} \ldots
| 0 \rangle_m^{\otimes n} \rightarrow | x \rangle_0 | f(x)
\rangle_1 \ldots | f^{(m)}(x) \rangle_m \label{eq17}.
\end{equation}
Our next step is to  perform the iteration which is inversed to
the one that  is defined by operator $ A_{m-1} $, where
\begin{equation}
A_{m-1}(A_{m-1})^{-1}=I.
\end{equation}
Applying of such an operator to the registers system after $m $
iterations (\ref{eq17}) can be written as
\begin{eqnarray}
(A_{m-1})^{-1}:   | x \rangle_0 | f(x) \rangle_1 \ldots |
f^{(m-1)}(x) \rangle_{m-1} | f^{(m)}(x) \rangle_m \rightarrow
\nonumber \\ \rightarrow | x \rangle_0 | f(x) \rangle_1 \ldots | 0
\rangle_{m-1}^{\otimes n} | f^{(m)}(x) \rangle_m \label{eq18}.
\end{eqnarray}
A set of inversed transformations can be written as
\begin{eqnarray}
(A_{1})^{-1} \ldots (A_{m-2})^{-1}(A_{m-1})^{-1}:   | x \rangle_0
| f(x) \rangle_1 \ldots | f^{(m-1)}(x) \rangle_{m-1} | f^{(m)}(x)
\rangle_m \rightarrow \nonumber \\ \rightarrow | x \rangle_0 | 0
\rangle_{1}^{\otimes n}  \ldots | 0 \rangle_{m-1}^{\otimes n} |
f^{(m)}(x) \rangle_m \label{eq19}.
\end{eqnarray}
As a result of applying the operators of inversed evolutions, the
registers of additional qubits will be in initial states and can
be removed without the affection on other qubits. General
evolution of CA can be described by operator
\begin{equation}
U_{CA}=(A_{1})^{-1} \ldots (A_{m-2})^{-1}(A_{m-1})^{-1} A_{m}
A_{m-1}  \ldots A_{1}.
\end{equation}
The resulted evolution of quantum CA can be written as
\begin{equation}
U_{CA} : | x \rangle | 0 \rangle ^ {\otimes n} \rightarrow | x
\rangle | f^{(m)}(x) \rangle .
 \label{eq21}
\end{equation}

Let us consider the steps of implementation of quantum cellular
automata:
\begin{enumerate}
\item At the first step   the CA register in the basic state is
initialized
\begin{equation}
 |x_1,x_2,\ldots x_n\rangle_n \rightarrow |0_1,0_2,\ldots 0_n\rangle_n
 \label{eq22}
\end{equation}
\item Let us apply the Hadamar operator  to each qubit of CA
register. This operator is given by matrix
\begin{equation}
H =\frac{1}{\sqrt{2}} \left( \begin{array}{cc}
1 & 1   \\
1 & -1    \\
\end{array} \right)
\label{eq23}.
\end{equation}
\item As a result we obtain the following superposition
\begin{equation}
H^{\otimes n}|0_1,0_2,\ldots 0_n\rangle_n= \frac{1}{\sqrt{2^n}}
\sum_{x=1}^{x=2^n}{|x \rangle}
\end{equation}
where $|x\rangle$  denotes   computational basis states
$|x_1,x_2,\ldots x_n\rangle_n$

The    Hadamar transformation generates the superposition of all
possible $2^n$ states with the equal amplitudes.

Let us apply unitary transformation (\ref{eq21})
\begin{equation}
|\Psi_{CA}\rangle=U_{CA} \left( H^{\otimes n} |0\rangle^{\otimes
n} \right) |0\rangle^{\otimes n} = \frac{1}{\sqrt{2^n}}
\sum_{x=1}^{2^n}{|x\rangle|f^{(m)}(x)}\rangle\label{eq23}
\end{equation}
Superposition $|\Psi_{CA}\rangle$ includes all $2^n$ results of
evolutions initial states of CA with dimension n. With one quantum
evolution of CA, all possible CA evolutions for initial states
with defined CA updating rules are performed.

\end{enumerate}

\section*{Analysis of Cellular Automata Evolution}

Let us consider the ability to amplify the amplitudes of searched
states with using the ideas of Grover's algorithm which is being
being used for searching files in the quantum database
\cite{stat4,stat5,stat6}. The difference between  this task and
Grover's algorithm is that unknown states are  not being searched
in this task, these searched states are known and we need to get
the answer that such states exist in CA evolution. Let us
introduce an additional qubit - ancilla which will be controlled
by n-qubit Toffoli gate where n-qubit of CA state are controlling
the ancilla state in the Toffoli gate. This gate can be described
by unitary matrix
\begin{equation}
T = \left( \begin{array}{cccc}
1 & \ldots & \ldots &  0   \\
\ldots & \ldots & \ldots &  \ldots \\
\ldots & \ldots & 0 & 1 \\
0 & \ldots & 1 & 0   \\
\end{array} \right)
\label{eq27}.
\end{equation}

Let us consider the unitary operator which is defined as tensor
product of one qubits operators
\begin{equation}
 S_T=\bigotimes_{i=1}^{n}S_i
 \label{eq28}
\end{equation}
where
\begin{equation}
 S_i= \left\{  \begin{array}{ll}
I, & \textrm{ïªé® $e_i=1$} \\
X, &  \textrm{ïªé® $e_i=0$}
\end{array} \right.
\label{eq29}
\end{equation}
The operator  $S_T$  flips searched states $|e_1,e_2,\ldots
e_n\rangle $ into the state $|1_1,1_2,\ldots 1_n\rangle$. It is
necessary for  implementation of anclilla inversion   for searched
states by Toffoli transformation (\ref{eq27})  Apply hadamar
operator to ancilla  $|z\rangle $  in prepared state $|1\rangle $
\begin{equation}
 |z\rangle = H |1\rangle = \frac{1}{\sqrt{2}}( |0\rangle -|1\rangle)
 \label{eq30}
\end{equation}

Let consider an operator

\begin{equation}
U_T =
 (I^{\otimes n} \otimes S_T \otimes I )( I^{\otimes n}
 \otimes T_n )
 ( I^{\otimes n} \otimes S_T \otimes I)
 \label{eq31}
\end{equation}

Let apply this operator to the system of qubits registers $
|x\rangle |f^{(m)}(x)\rangle |z\rangle $. the first group of
operators on the right side which is delimited by brackets
performs flipping into the state $|1_1,1_2,\ldots 1_n\rangle$ of
searched finish states, the second group performs ancilla $
|z\rangle $ inversion for searched states and the third group
returns states which are changed by the first group into the state
before applying the operator. Additional controlled qubit  stays
in the state (\ref{eq30}). The result of   the operator
(\ref{eq31}) acting is
\begin{equation}
U_T(|x\rangle |f^{(m)}(x)\rangle |z\rangle) = \frac{1}{\sqrt{2^n}}
\left( \sum_{x \notin X_q} |x\rangle |f^{(m)}(x)\rangle - \sum_{x
\in X_q} |x\rangle |q\rangle \right) \otimes |z\rangle
 \label{eq32}
\end{equation}
where $X_q$ - is set of initial states $|x \rangle$, which leads
to the searched states $|q\rangle$ in the CA evolutional process
after $m$ CA iterations. Ancilla $ |z\rangle $ in a new basis
leaves in the unchanged state, but in the superpositions states
the inversion of amplitudes signs for subsystem $\sum_{x \in X_q}
|x\rangle |q\rangle$ appears. It is caused by the transformation
of ancilla $ |z\rangle $ to new state  (\ref{eq30}) before
applying the operator (\ref{eq32}).

Let us consider the inversion operator from Grover's algorithm
\begin{equation}
U_G = 2 | \Psi_c \rangle \langle \Psi_c | - I
 \label{eq33}
\end{equation}
where
\begin{equation}
| \Psi_c \rangle = H^{\otimes n}|0\rangle^{\otimes n}=
\frac{1}{\sqrt{2^n}} \sum_{i=1}^{i=2^n}{|i \rangle}
\end{equation}
This operator will be used for  the states  $\sum_{x \in X_q}
|x\rangle |q\rangle $ amplitudes amplification. The operator $U_G$
reflect any vector around the axis defined by vector $|\Psi_c
\rangle$. The operator $U_G$ can be decomposed by one-qubit
operators
\begin{equation}
U_G = H^{\otimes n} (2 |0\rangle \langle 0 | -I) H^{\otimes n}
\label{eq34}
\end{equation}
The operator $U_G$ also has been called as the operator of
inversion about the average. The transformation (\ref{eq33}) can
also be defined as
\begin{equation}
U_G  : \sum_{i}{a_i |i\rangle} \rightarrow  \sum_{i}{(2A-a_i)
|i\rangle}  \label{eq35}
\end{equation}
where $A$ - means the average value of $a_i$ amplitudes.

\begin{figure}[h]
\centering
\includegraphics[width=0.4\textwidth]{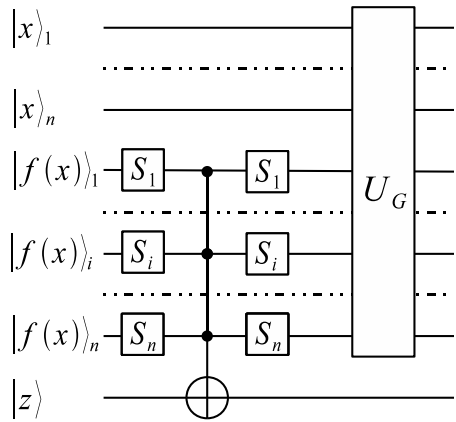}
\caption {{\small  Quantum circuit implementation of the
transformation $U_G U_T$ }} \label{fig2}
\end{figure}

If searched state $\sum_{x \in X_q} |x\rangle |q\rangle $ can be
realized only once from one initial CA state, then its amplitude
is
\begin{equation}
\beta = \frac{1}{\sqrt{2^n}}
\end{equation}
It can be shown that performing of CA iteration
\begin{equation}
U_G U_T(|x\rangle |f^{(m)}(x)\rangle |z\rangle) \label{eq37}
\end{equation}
we obtain the amplitude amplification by 3 times that match the
implementation of inversion iteration in the Grover's algorithm
\cite{stat4,stat5,stat6}. The transformation $U_G U_T$ can be
described by quantum circuit in fig.\ref{fig2}.

 Let us consider the number of iterations necessary for
enough amplification of searched states amplitudes. If only one
initial state $|x\rangle$ of cellular automata leads to the
searched finish state $|q\rangle$,  then using the ideas similar
to Grover's algorithms \cite{stat4,stat5,stat6} we can find the
optimal number of unitary transform $U_G U_T $ .
\begin{equation}
k \approx \frac{\pi}{4} \sqrt{N}, N=2^n  \label{eq38}
\end{equation}
It follows from this result that algorithm complexity is
$O(\sqrt{N})$. In comparision with analogical classical algorithm
$O(N)$  it means polinomial computational speedup. If instead 1
matching initial state there are  $ l $ matching states which lead
to the searched finish state, then
\begin{equation}
k \approx \frac{\pi}{4} \sqrt{\frac{N}{l}},  \label{eq39}
\end{equation}

if $l$ is unknown, then Grover's algorithm could be run several
times when
\begin{equation}
L = 1,2,4,8\ldots
\end{equation}
It can be shown, that for such series calculation the complexity
will be still $O(\sqrt{N})$.

\section*{Conclusion}
It is shown that irreversible classical cellular automata can be
performed by quantum algorithm using additional ancilla registers.
The algorithm for cellular automata states analysis has been
proposed. This algorithm is based on the elements of Grover's
algorithm - the inversion of amplitude of searched states and
unitary transform of inversion about the average. The  inversion
of searched states amplitudes can be performed by quantum Toffoli
gate.

\end{document}